# A Catalog of Type II radio bursts observed by Wind/WAVES and their Statistical Properties

Nat Gopalswamy[1], Pertti Mäkelä[1,2], and Seiji Yashiro[1,2]

[1]NASA Goddard Space Flight Center, Greenbelt, MD 20771, USA

[2]The Catholic University of America, Washington DC 20064, USA

e-mail: nat.gopalswamy@nasa.gov

Abstract:

Solar type II radio bursts are the signature of particle acceleration by shock waves in the solar corona and interplanetary medium. The shocks originate in solar eruptions involving coronal mass ejections (CMEs) moving at super-Alfvenic speeds. Type II bursts occur at frequencies ranging from hundreds of MHz to tens of kHz, which correspond to plasma frequencies prevailing in the inner heliosphere from the base of the solar corona to the vicinity of Earth. Type II radio bursts occurring at frequencies below the ionospheric cutoff are of particular importance, because they are due to very energetic CMEs that can disturb a large volume of the heliosphere. The underlying shocks accelerate not only electrons that produce the type II bursts, but also protons and heavy ions that have serious implications for space weather. The type II radio burst catalog (https://cdaw.gsfc.nasa.gov/CME_list/radio/waves_type2.html) presented here provides detailed information on the bursts observed by the Radio and Plasma Wave Experiment (WAVES) on board the Wind Spacecraft. The catalog is enhanced by compiling the associated flares, CMEs, solar energetic particle (SEP) events including their basic properties. We also present the statistical properties of the radio bursts and the associated phenomena, including solar-cycle variation of the occurrence rate of the type II bursts.

## 1. Introduction

Type II radio bursts were discovered by Payne-Scott, Yabsley, and Bolton (1947) as radio emission progressively starting at lower frequencies. This drifting nature was already recognized by these authors as an indication of mass motion away from the Sun in the form of filament eruptions. The drifting bursts were classified type II radio bursts by Wild and McCready (1957). The plasma emission mechanism (Ginzburg and Zhelezniakov, 1958) is thought to be responsible for type II bursts. There is a wealth of literature on type II radio bursts from ground-based spectral and imaging observations as reviewed by Nelson and Melrose (1985). When space radio instruments became available, type II bursts were observed at frequencies below the ionospheric cutoff. Malitson, Fainberg, and Stone (1973a) reported a type II burst at 2.2 MHz and slowly drifting down to lower frequencies over several hours, and down to local plasma frequency at 1 AU (Malitson, Fainberg, and Stone, 1973b). Observations by the Voyager Radio Astronomy experiments have shown that type II bursts in the interplanetary (IP) medium are rather common (Boischot et al., 1980). Shocks from the Sun were thought to be responsible for the radio bursts as was initially suggested by Uchida (1960) for coronal type II bursts. The ISEE-3 spacecraft observed many IP type II bursts in the frequency range 30 kHz to 2 MHz (Cane, Sheeley, and Howard, 1987; Lengyel-Frey and Stone, 1989) and the shock driver was identified as coronal mass ejections (CMEs). All these IP type II observations were made at frequencies below 2 MHz. The Radio and Plasma Wave Experiment (WAVES, Bougeret et al. 1995) on board the Wind spacecraft launched in 1994 was able to fill the frequency gap in type II observations between 2 and 14 MHz. Wind/WAVES observations enabled tracking shocks from the corona to 1 AU (Reiner et al. 1999).

The idea that shocks responsible for type II bursts are driven by magnetic structures from the Sun (Fokker 1963) preceded the discovery of CMEs in white light (Tousey 1973). Stewart et al. (1974) confirmed that the shock underlying the type II burst is indeed driven by the white-light transient. The shocks indicated by type II bursts are important players in the solar terrestrial system: they cause SEP events (Kahler et al. 1978), sudden commencements (as originally envisioned by Gold, 1955), compress the magnetosphere, produce energetic storm particle events inside the magnetosphere, and start geomagnetically induced currents. While only tens of IP type II bursts and the associated CMEs were known in the pre-Wind era, the launch of the Solar and Heliospheric Observatory (SOHO) resulted in hundreds of type II burst and CME pairs observed simultaneously with a better frequency coverage (e.g., Gopalswamy et al. 2001; 2005a). This led to many important discoveries on the energetic phenomena in the coronal region (2 to 10 solar radii, Rs) not accessed by pre-Wind radio instruments (see, e.g., Gopalswamy 2004).

Recognizing the importance of type II bursts in identifying shock-driving CMEs in assessing the geoeffectiveness of CMEs and their ability to cause SEP events, a catalog of type II bursts observed by Wind/WAVES has been created at the CDAW Data Center of NASA's Goddard Space Flight Center (NASA/GSFC). In this paper, we describe the catalog including the added features that enable researchers to study the underlying shocks from their origin at solar source regions to their consequences in the heliosphere. In addition, we present the statistical properties of DH type II bursts, the associated CMEs and SEP events, taking advantage of the extended data available over two solar cycles.

## 2. The Wind/WAVES Type II Burst Catalog

Type II bursts are due to shocks driven by CMEs. These CMEs can occur anywhere on the disk and even from the backside. These CMEs are generally accompanied by large solar flares, which help us identify the location on the Sun from which the underlying eruption occurs. Type II bursts are also good indicators of large SEP events because the same shock accelerates electrons and ions. Therefore, we decided to develop a comprehensive catalog that compiles all the related phenomena in one place with type II radio bursts in the decameter-hectometric (DH) domain as the primary data obtained by Wind/WAVES. The WAVES radio instrument consists of two radio receivers called Radio Receiver Band 1 (RAD1, 1040–20 kHz) and Band 2 (RAD2, 13.825–1.075 MHz). These correspond to wavelength ranges 288.3 m–15 km and 21.7–279 m, respectively. Thus, RAD2 covers the decameter-hectometer band, while RAD1 covers the hectometer-kilometer band.

Type II bursts are identified as slowly drifting features in the radio dynamic spectra, which are frequency-time plots of radio intensity. An early effort to list type II bursts observed by Wind/WAVES was done by M. L. Kaiser until March 2015 (https://solar-radio.gsfc.nasa.gov/wind/data_products.html). We updated this list and added flare and CME information in order to understand the solar source and heliospheric consequences of the type II bursts. The CME data are from the Large Angle and Spectrometric Coronagraph (LASCO, Brueckner et al. 1995) on board the Solar and Heliospheric Observatory (SOHO, Domingo, Fleck, and Poland, 1995) mission. The CMEs in this catalog are called radio-loud CMEs because of their ability to produce type II radio bursts (Gopalswamy et al. 2008a). The flare locations are derived from the online Solar Geophysical Data listing or from inner coronal images such as Yohkoh's Soft X-ray Telescope (SXT, Tsuneta et al. 1991) and SOHO's Extreme-ultraviolet imaging telescope (EIT, Delaboudinière et al. 1995). Some solar source locations have also been obtained from Solarsoft Latest Events Archive after October 1, 2002: http://www.lmsal.com/solarsoft/latest_events_archive.html. Information on flare sizes and SEP events are obtained from NOAA's GOES satellite data.

```
=================================================================================================
              DH Type II                    Flare                    CME
         ---------------------------     -----------------    ---------------------------         Plots
Start         End           Frequency     Loc  NOAA  Imp      Date  Time  CPA  Width  Spd
(1)           (2)     (3)   (4)  (5) (6)  (7)   (8)   (9)     (10)  (11)  (12) (13)  (14)          (15)
=================================================================================================
1997/04/01 14:00 04/01 14:15  8000  4000  S25E16  8026 M1.3   04/01 15:18   74    79   312         PHTX
1997/04/07 14:30 04/07 17:30 11000  1000  S28E19  8027 C6.8   04/07 14:27 Halo   360   878         PHTX
1997/05/12 05:15 05/14 16:00 12000    80  N21W08  8038 C1.3   05/12 05:30 Halo   360   464         PHTX
1997/05/21 20:20 05/21 22:00  5000   500  N05W12  8040 M1.3   05/21 21:00  263   165   296         PHTX
1997/09/23 21:53 09/23 22:16  6000  2000  S29E25  8088 C1.4   09/23 22:02  133   155   712         PHTX
1997/11/03 05:15 11/03 12:00 14000   250  S20W13  8100 C8.6   11/03 05:28  240   109   227         PHTX
1997/11/03 10:30 11/03 11:30 14000  5000  S16W21  8100 M4.2   11/03 11:11  233   122   352         PHTX
1997/11/04 06:00 11/05 04:30 14000   100  S14W33  8100 X2.1   11/04 06:10 Halo   360   785         PHTX
1997/11/06 12:20 11/07 08:30 14000   100  S18W63  8100 X9.4   11/06 12:10 Halo   360  1556         PHTX
1997/11/27 13:30 11/27 14:00 14000  7000  N17E63  8113 X2.6   11/27 13:56   98    91   441         PHTX
1997/12/12 22:45 12/12 23:20 14000  8000  N25W52  8116 B9.4   12/13 00:26  278    73   191         PHTX
1998/01/25 15:03 01/25 15:18 14000 10000  N21E25  8141 C1.1   01/25 15:26 Halo   360   693         PHTX
1998/03/29 03:40 03/29 03:52 14000  7000    SW90  ----- ----  03/29 03:48 Halo   360  1397         PHTX
1998/04/20 10:25 04/22 06:00 10000    35  S22W90  8194 M1.4   04/20 10:07  284   165  1863         PHTX
1998/04/23 06:00 04/23 15:30 14000   200  S17E90  8210 X1.2   04/23 05:55 Halo   360  1691         PHTX
1998/04/24 09:17 04/24 09:25  4700  2600  S10E90  8210 C8.9   04/24 08:55  100    84  1184         PHTX uncertain
1998/04/27 09:20 04/27 10:00 10000  1000  S16E50  8210 X1.0   04/27 08:56 Halo   360  1385         PHTX
1998/04/29 16:30 04/29 17:00 10000  2000  S18E20  8210 M6.8   04/29 16:58 Halo   360  1374         PHTX
1998/05/02 14:25 05/02 14:50  5000  3000  S15W15  8210 X1.1   05/02 14:06 Halo   360   938         PHTX
1998/05/06 08:25 05/06 08:35 14000  5000  S11W65  8210 X2.7   05/06 08:29  309   190  1099         PHTX
1998/05/09 03:35 05/09 10:00  9000   400  S14W89  8210 M7.7   05/09 03:35  262   178  2331         PHTX
1998/05/11 21:40 05/11 22:00 10000  1000  N32W90  8214 B6.6   05/11 21:55  208  >301   830         PHTX
1998/05/19 10:00 05/19 11:30 14000  3000  N29W46  8222 B7.9   05/19 10:27  268   139   801         PHTX
1998/05/27 13:30 05/27 14:20  4000  1000  N19W62  8226 C7.5   05/27 13:45  175   268   878         PHTX
1998/06/11 10:15 06/11 10:20  8000  4000  N16E86  8243 M1.4   06/11 10:28  123   177  1223         PHTX
1998/06/16 18:20 06/17 21:00 12000    50  S22W90  8232 M1.0   06/16 18:27  341   281  1484         PHTX
1998/06/20 19:39 06/20 20:00  2600  1800    Back  ----- ----  06/20 18:20 Halo   360   964         PHTX uncertain
1998/06/22 07:15 06/22 09:20  6000  2000  N16W46  8243 C2.9   06/22 05:01  265    59   206         PHTX
```

Figure 1. A screenshot of the DH type II burst catalog showing information on the type II bursts and on the associated flares and CMEs. Entries highlighted in blue indicate that they are linked to movies and plots. Comments on some events are provided after the last column.

One of the key features of this catalog is the movies and plots included for each type II burst and the associated phenomena. These enable the user to make the association among various phenomena and to make further measurements online using the tools provided. Additional information can be obtained from the main catalog (https://cdaw.gsfc.nasa.gov/CME_list) for events listed in the type II catalog. The JavaScript movies consist of a series of images from SOHO/LASCO, Wind/WAVES dynamic spectra, and GOES soft X-ray light curves. All these images and plots can be downloaded.

## 2.1 Explanation of Catalog Entries

Figure 1 shows a screenshot of the catalog. The catalog is arranged in a tabular format, each row devoted to one type II burst observed by Wind/WAVES. After the launch of the Solar Terrestrial Relations Observatory (STEREO, Kaiser et al. 2008) in October 2006, some entries, especially the start and end times and frequencies of the bursts are determined using both Wind and STEREO/WAVES (Bougeret et al. 2008) observations. A detailed explanation of the catalog entries is as follows:

Column 1: Starting date of the type II burst (yyyy/mm/dd format)
Column 2: Starting time (UT) of the type II burst (hh:mm format)
Column 3: Ending date of the type II burst (mm/dd format; year in Column 1 applies)
Column 4: Ending time of the Type II burst (hh:mm format)
Column 5: Starting frequency of type II burst (kHz). An entry "????" indicates that the starting frequency is not determined. The starting frequency is linked to JavaScript movies that can be used to view the CMEs and the type II bursts together using the c2rdif_waves.html movies linked to the starting frequency.
Column 6: Ending frequency of type II burst (kHz). An entry "????" indicates that the ending frequency is not determined. The c3rdif_waves.html movies are linked to the ending frequencies.
Column 7: Solar source location (Loc) of the associated eruption in heliographic coordinates. E.g., S25E16 means the latitude is 25º south and 16º east (source located in the southeast quadrant of the Sun). N denotes northern latitudes and W denotes western longitudes. Entries like SW90 indicate that the source information is not complete, but we can say that the eruption occurs on the west limb but at southern latitudes; if such entries have a subscript b (e.g., NE90b) it means that the source is behind the particular limb. This information is usually gathered from SOHO/EIT difference images, which show dimming above the limb in question. Completely backside events with no information on the source location are marked as "back".

Column 8: NOAA active region number. If the active region number is not available or if the source region is not an active region, the entry is "----". Filament regions are denoted by "FILA" or "DSF" for disappearing solar filament.

Column 9: Soft X-ray flare importance (Imp). GOES soft X-ray flare size (peak flux in the 1-8 Å channel) is listed. $X1.0 = 1.0 \times 10^{-4}$ W m$^{-2}$ and the letters M, C, B, and A denote flares of progressively lower class in steps of an order of magnitude. "----" indicates that the soft X-ray flux is not available.

Column 10: Date of the associated CME (mm/dd format, Year in Column 1 applies). Details of the CME are also available in the main CME catalog (https://cdaw.gsfc.nasa.gov/CME_list). Lack of SOHO observations are noted as "LASCO DATA GAP". Other reasons are also noted if there are no CME parameters measured. The CMEs and the GOES flare light curves for a given type II burst can be viewed from the JavaScript movies linked to the CME date.

Column 11: Appearance Time of the associated CME (hh:mm format).

Column 12: Central position angle (CPA, degrees) for non-halo CMEs. CPA is meaningful only for non-halo CMEs. For halo CMEs, the entry is "Halo". For halo CMEs, the height-time measurements are made at a position angle where the halo appears to move the fastest. This is known as the measurement position angle (MPA) and can be found in the main catalog (http://cdaw.gsfc.nasa.gov/CME_List).

Column 13: CME width in the sky plane (degrees). Width = 360 means the CME is a full halo. Widths prefixed by ">" indicates that the reported width is a lower limit.

Column 14: CME speed in the sky plane (km/s). This speed is likely to be the true speed only for CMEs originating from close to the limb. For other CMEs, the sky-plane speed is generally smaller than the true speed due to projection effects. The height-time plots of the CMEs (with linear and quadratic fits to the data points superposed) are linked to the CME speed.

Column 15: Link to three-day overview plots of solar energetic particle events (protons in the >10, >50 and >100 MeV GOES energy channels), CME height-time, and X-ray flare (PHTX) plots.

Notes are added at the end of a row to comment on any peculiarities. For example, the identification of the type II burst may be uncertain or there may be a data gap in CME observations.

## 2.2 Illustrative Examples

Figure 2 shows an event that occurred on 2005 January 15 toward the end of the day. The prominent features on the radio dynamic spectrum are the noise storm, type III burst, type IV burst, and the type II burst, which is the main feature listed in the catalog. All the radio emissions are related to the eruption one way or the other. The noise storm is actually ongoing before the eruption but gets disrupted by the eruption. One can see that there is no noise storm emission after the type III burst. It is thought that the noise storms are produced by low-energy electrons accelerated in the field lines overlying the active region that get disrupted during the eruption. The type III bursts are produced by electrons accelerated at the flare site and propagating along open field lines in the IP medium. The type IV bursts, on the other hand, are produced by flare electrons trapped in post-eruption arcades that are anchored to the Sun. The type IV burst is actually an extension of the higher frequency counterparts observed at metric wavelengths. The type II burst is thought to originate in the shock front, which is a highly extended around the CME flux rope. It is generally believed that the type II bursts originate from the nose region of the shock, where the shock is strongest. At the time of the LASCO image, the radio emission is in progress at a frequency of ~1 MHz; the shock structure is at a heliocentric distance of 24.95 Rs in sky-plane projection. The evolution of the CME/shock structure and the type II burst can be observed from the movie, [https://cdaw.gsfc.nasa.gov/movie/make_javamovie.php?date=20050115&img1=lasc3rdf&img2=wwaves](https://cdaw.gsfc.nasa.gov/movie/make_javamovie.php?date=20050115&img1=lasc3rdf&img2=wwaves), made available on the web site. This movie has LASCO/C3 images and the radio dynamic spectrum. The LASCO/C2 images have SOHO/EIT difference image taken around the same time superposed, so we can identify the solar source of the eruption.

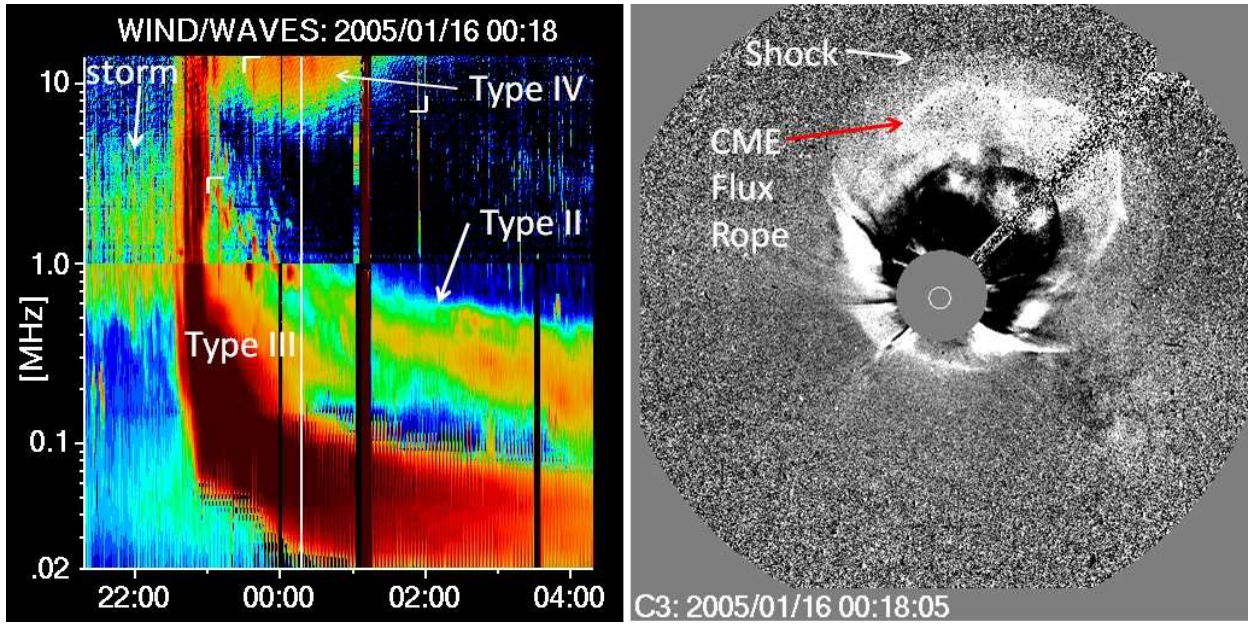

Figure 2. Wind/WAVES dynamic spectrum (left) and the associated white-light CME observed by SOHO/LASCO (right). The dynamic spectrum shows four different types of radio emission: type III storm, which consists of a series of short duration bursts forming a band between 0.4 and 5 MHz, intense type III burst starting around 22:30 UT and lasting until 23:00 UT, a type IV burst starting at the end of the type III burst and ending around 01:45 UT (next day) between the top of the dynamic spectrum and 6 MHz, and the type II burst starting around 23:00 UT and lasting for many hours. The type II burst starts at a frequency of 5 MHz and extends to kHz frequencies. From the SOHO/LASCO image at 00:18:05 UT, a previous image has been subtracted to show the changes. The dark part of the image actually is the position of the CME in the previous image that has been subtracted. The difference image shows the flux rope structure (the bright feature) and the diffuse shock structure ahead. The shock itself is too thin to image but is supposed to be located at the leading edge of the diffuse sheath region. The vertical white line on the dynamic spectrum marks the time of the LASCO/C2 image. The type II burst above 1 MHz is weak and fragmented mainly because of the difference in sensitivity of the two receivers RAD1 and RAD2.

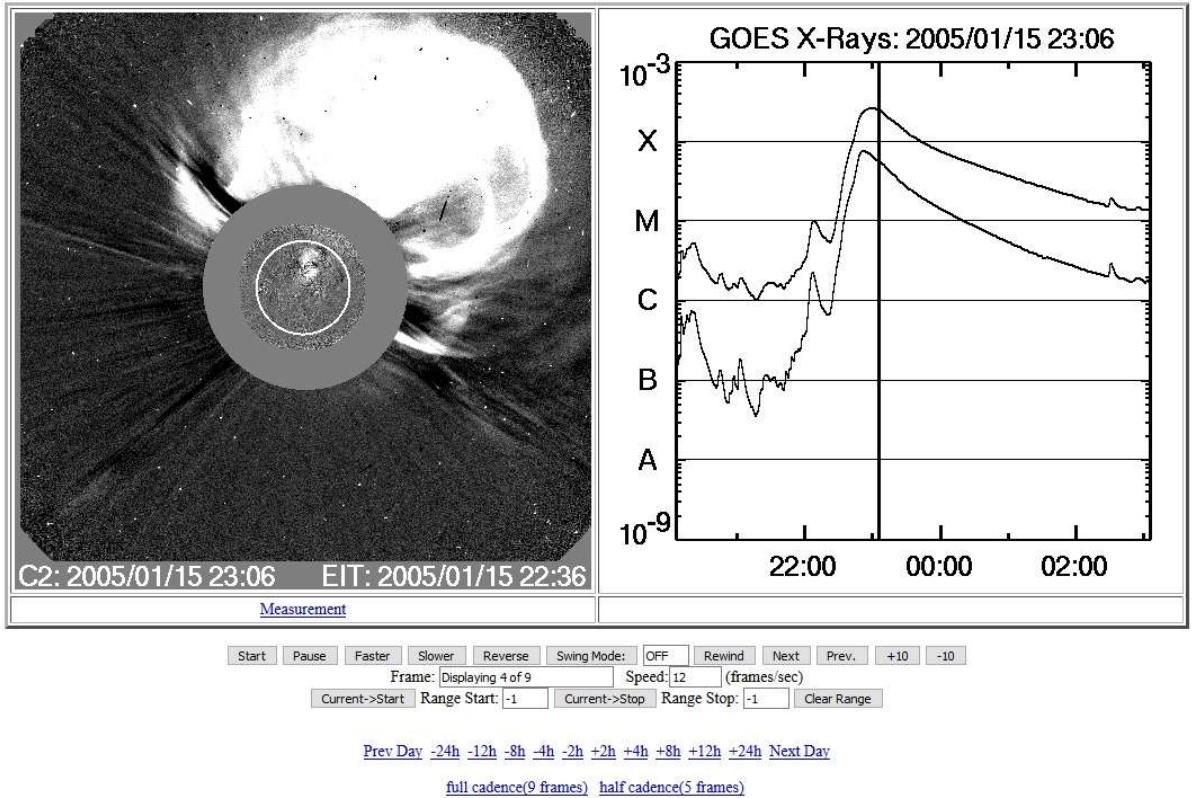

Figure 3. The actual appearance of a frame in the JavaScript movie linked to the starting frequency of the 2005 January 15 DH type II burst at 23:00 UT in the catalog. The link is: https://cdaw.gsfc.nasa.gov/movie/make_javamovie.php?stime=20050115_2144&etime=20050116_0108&img1=lasc2rdf&title=20050115.230650.p323s;V=2861km/s.

Figure 3 shows a snapshot of the JavaScript movie showing the LASCO/C2 images combined with GOES soft X-ray light curves. The CME identifier appears at the top as "20050115.230650.p323s;V=2861km/s" which means the CME first appeared in LASCO/C2 field of view (FOV) on 2005 January 15 at 23:06:50 UT at position angle 323º and had an average speed of 2861 km/s. The movie combines LASCO/C2 difference images with superposed EIT difference images with GOES soft X-ray light curves in the 1-8 Å (upper) and 0.4-5 Å (lower) energy channels. The vertical dark line indicates the time of the LASCO/C2 image. In the frame shown, the time of the LASCO/C2 image is 23:06 UT, which is also the time of first appearance in the C2 FOV. The shock part of the CME has already left the FOV, and we see only the flux rope part (the bright feature); the shock can be seen only on the sides of the flux rope. The superposed EIT difference image corresponds to 22:36 UT (the nearest available) and shows solar source location as a disturbance on the solar disk. The link "Measurement" provided below the CME frame opens a new window with an online tool to measure the height-time history of any feature that is seen in the images. Below the movie frame, a number of buttons are provided to play the movie in various ways. The time window of the movie can also be extended to earlier and later times by clicking on the links between "Prev day" and "Next Day". Additional links are also provided to play the movie with full cadence or half cadence images.

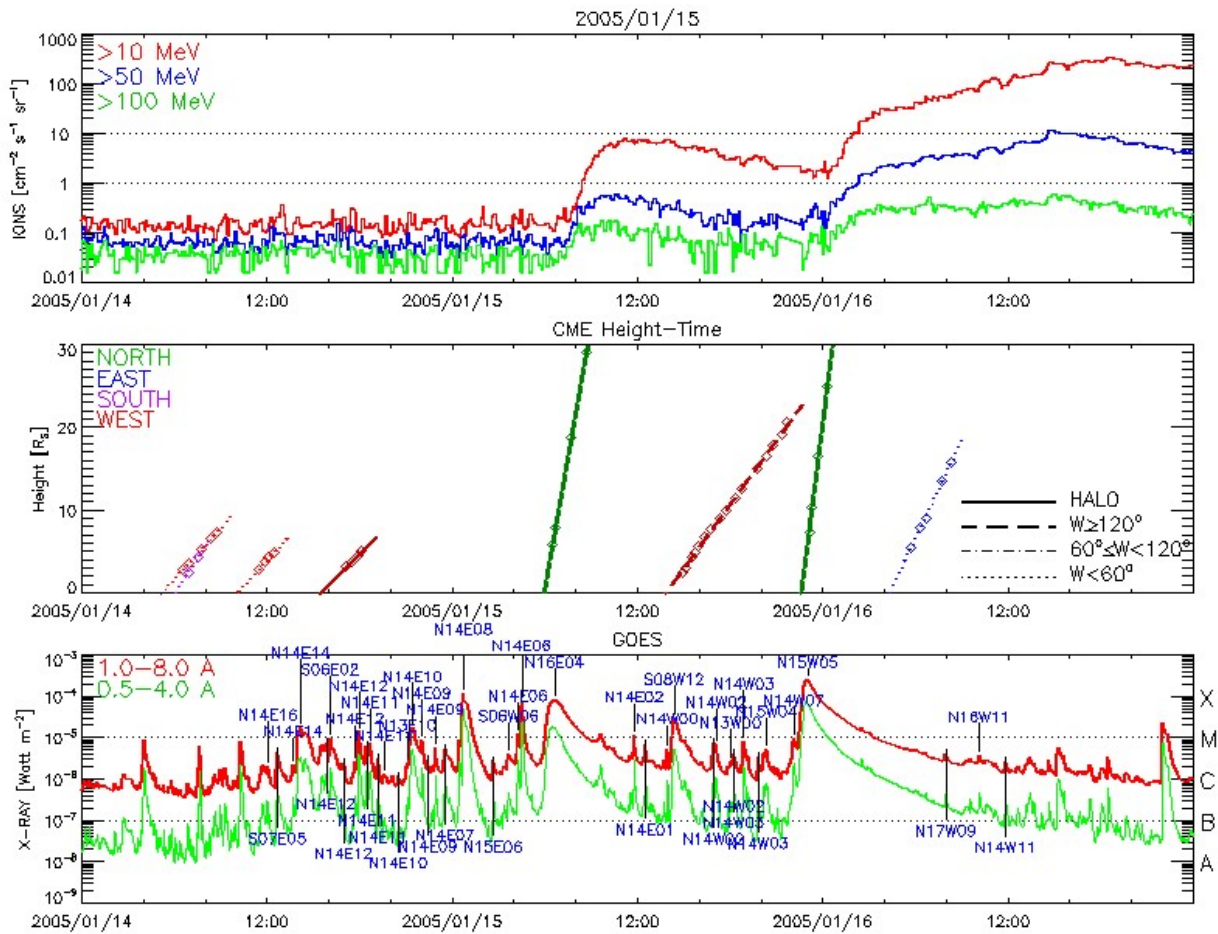

Figure 4. The 3-day "PHTX" plot included in the catalog providing information on SEPs, CMEs, and flares. (top) Intensity of energetic protons detected by the GOES satellite given in three integral energy channels, >10 MeV, > 50 MeV, and >100 MeV. (middle) Height-time plots where CMEs are distinguished by their widths W with different line styles (HALO for W=360º, W ≥120º, 60º≤W<120º, and W<60º) and predominant directions (color coded). (bottom) GOES soft X-ray light curves in two energy channels (1-8 Å and 0.5 to 4 Å) showing the flares and their heliographic coordinates.

Figure 4 shows the "PHTX" plot that is linked to the entry of the 2005 January 15 DH type II burst at 23:00 UT. There are two increases in proton intensity in the top panel corresponding to the two ultrafast CMEs on 2005 January 15 seen in the middle panel (green height-time plots). If the >10 MeV plot in the top panel exceeds the 10 pfu line, it is deemed as a large SEP event. According to this definition, the second rise is a large SEP event, while the first rise is not. The corresponding CMEs are halos, have similar speeds (2049 and 2861 km/s in the sky plane), and have a homologous appearance in the coronagraph FOV and in Wind/WAVES dynamic spectra. The second CME is shown in Fig. 2. The bottom panel of Fig. 4 shows that there were lots of flares on this day and the flare associated with the CME of interest is the X-class flare originating from N15W05 (see also Fig. 3). The partial halo CME between the two halo CMEs originated from the southern hemisphere (S08W12) in association with an M-class flare. This CME is slow (~498 km/s) and is not associated with a DH type II burst or a particle event.

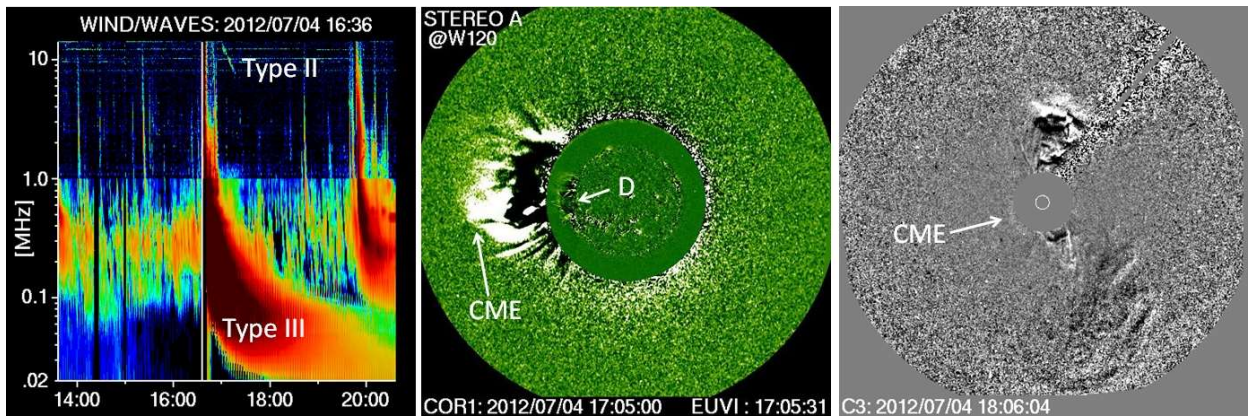

Figure 5. Wind/WAVES dynamic spectrum (left), STEREO-A/COR1 running difference image at 17:05 UT with a superposed EUVI difference image (middle), and LASCO/C3 difference image at 18:06 UT showing the halo CME (right). In the dynamic spectrum, the type II burst is the thin slanted feature between 17:00 UT and 17:20 UT. The type II burst is preceded by a type III burst. In the COR1 image, the CME leading edge is at a height of 3.5 Rs. The source was located just on the limb as can be seen from the EUVI disturbance marked D in the difference image at 17:05:31 UT.

Figure 5 shows a second example in which the type II burst is extremely weak, narrow band, and short-lived. The burst appears near 14 MHz and drifts down to about 7.5 MHz in about 20 min. The associated CME is listed in the catalog as a relatively slow halo CME (~660 km/s) originating from NOAA active region 11513 located at N14W34. The CME first appeared in LASCO/C2 FOV at 17:24 UT, at which time the DH type burst has already ended. The Atmospheric Imaging Assembly (AIA, Lemen et al. 2012) on board the Solar Dynamics Observatory (SDO, Pesnell, Thompson, and Chamberlin, 2012) imaged the eruption as a disturbance spreading from the source region. Fortunately, the source was located on the east limb of the Sun in the view of STEREO-Ahead (STA) spacecraft, which was ahead of Earth by 120°. The solar source at W34 in SOHO view corresponds to E86 in STA view. The height-time measurements using the COR1 coronagraph images obtained by the Sun Earth Connection Coronal and Heliospheric Investigation (SECCHI, Howard et al. 2008) on board STEREO yield a speed of 1276 km/s at 17:03 (using a second order polynomial fit to the five height-time data points) with the CME leading edge height at 3.27 Rs. We can cross check this speed with the speed derived from the drift rate of the DH type II burst. Again, using the measurement tool available online, we measure the drift rate of the DH type II burst as $6.36 \times 10^{-3}$ MHz/s. At 17:03 UT, the emission frequency is 11 MHz and the shock height $r = 3.27$ Rs. The shock speed is related to the drift rate according to:

$$V = 2L(1/f)(df/dt) \quad (1)$$

where $L$ is the density ($n$) scale height given by $L = [(1/n)|dn/dr|]^{-1}$. At a height of 3.27 Rs, the density varies as $r^{-2}$, so we get $L = r/2$. With these parameters, we get $V = 1316$ km/s, in agreement with the leading-edge speed (1276 km/s) obtained from COR1 height-time measurements. Here we tacitly assumed that the DH type II emission is at the fundamental plasma frequency. In fact, a composite dynamic spectrum constructed from Wind/WAVES and ground-based Radio Solar Telescope Network (RSTN) data indicate that the DH type II burst is the extension of the fundamental component of the metric type II burst that showed both fundamental and harmonic structures (not shown here).

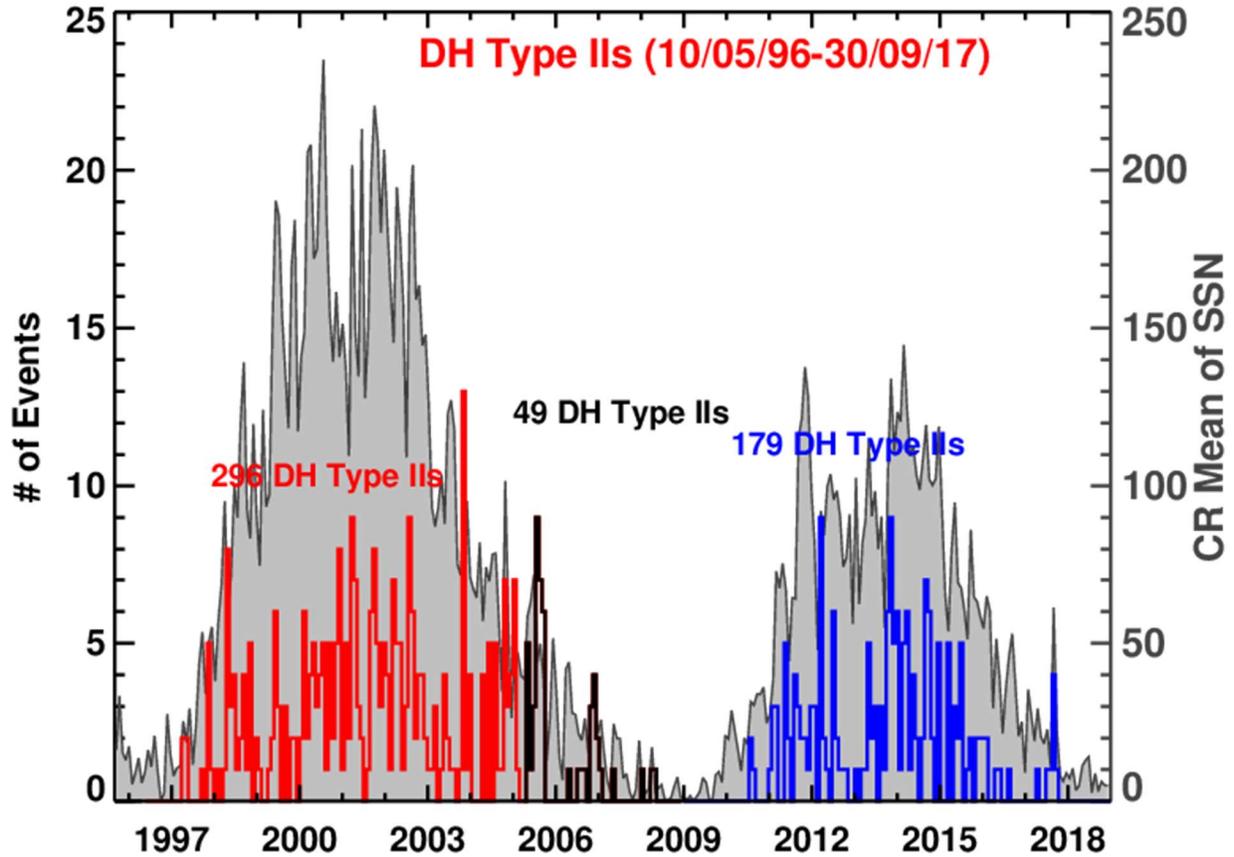

Figure 6. Occurrence rate of DH type II bursts summed over Carrington rotation (CR) periods from 1996 May 10 to 2017 September 30. This plot includes 524 DH type II bursts, which are grouped into three intervals: the red (296 bursts) and blue (179 bursts) bins correspond to the first 106 months in cycles 23 and 24, respectively. The dark bins (49 bursts) correspond to the rest of the period between the two cycles. The daily sunspot number (SSN), averaged over CR periods, is shown in gray for comparison. The version 2 international sunspot number data are from the World Data Center - Sunspot Index and Long-term Solar Observations (WDC-SILSO), Royal Observatory of Belgium, Brussels (http://sidc.be/silso/datafiles).

**2.3 Solar Cycle Variation**

Wind observations started in November 1994 during the deep minimum of solar cycle 22 before the start of cycle 23 in 1996. As of 2017 September 30, a total of 528 DH type II bursts have been identified by visual examination of the Wind/WAVES dynamic spectra. These include all type II bursts that have a starting frequency between 1 and 14 MHz (DH domain). This means, purely km type II bursts are not listed, but can be identified from daily dynamic spectra available in the SOHO/LASCO catalog (https://cdaw.gsfc.nasa.gov). In four cases, the spectral feature is not clear, so it was not possible to identify the starting and ending frequencies. These four events are not included in this study. The remaining 524 events are shown in Fig. 6 binned over Carrington (CR) rotation periods. Since the sunspot number (SSN) is taken as the primary index of solar activity, we have shown it (also binned over CR periods) for comparison. There is an overall good correspondence between SSN and DH type II rate, including the prominent double peak in SC 24.

**2.3.1 Inter-cycle variation**

Since DH type II bursts have been observed during solar cycles 23 and 24, it is possible to study the inter-cycle variation of the occurrence rates in the two cycles. Cycle 24 is the smallest cycle since the beginning of space age and is significantly weaker than cycle 23. Therefore, it is of interest to see if the occurrence rate of DH type II bursts shows similar variations. The nominal start and end dates of cycle 23 are 1996 May 10 and the end date is 2008 November 30. Solar cycle 24 started on 2008 December 1 and is progressing towards its minimum, which might occur over the next year or so. The number of months in cycle 24 from 2008 December 1 to 2017 September 30 is 106. The last DH type II burst occurred (as of this writing) occurred on 2017 September 17. The corresponding epoch in cycle 23 ended on 2005 March 9 (106 months from 1996 May 10). We compare the occurrence rate of type II bursts over the first 106 months in each of cycles 23 and 24.

Figure 6 shows that there are only 179 bursts in SC 24 compared to 296 in cycle 23 over the corresponding epoch. There are 49 more type II bursts until cycle 23 ended in 2008. The occurrence rate over the 106 months thus drops from 296 in cycle 23 to 179 in cycle 24 (by 40%). Over the first 106 months, the average SSN in cycles 23 and 24 are 107 and 60, respectively indicating a drop of 44%. Thus, the drop in the number of DH type II bursts is slightly smaller than that in SSN. The number of fast and wide (FW) CMEs (speed ≥900 km/s and width ≥60º) in the two cycles are 425(SC 23) and 251 (SC 24) representing a drop of 41%. For four months (July to September 1998 and January 1999), SOHO was temporarily disabled, so no CME data are available. Assuming that the occurrence rate of FW CMEs during the four months is the same as in the other 102 months, we estimate 17 FW CMEs during the 4 months.

This results in an estimated total of 442 FW CMEs over the first 106 months in cycle 23, indicating a drop of 43%. DH type II occurrence is thus smaller than that of FW CMEs and SSN. Moreover, the total number of FW CMEs (693) is only slightly larger than the number of DH type II bursts (524) as of this writing. During the first 106 months the number of regular CMEs (width $\geq 30^0$) are: are 6967 (cycle 23) and 8163 (cycle 24) as counted using the search engine in the SOHO/LASCO CME catalog (http://www.lmsal.com/solarsoft/www_getcme_list.html). As for FW CMEs, we accounted for the 4-month SOHO data gap in cycle 23 by assuming that the regular CMEs during the gap occurred with the same average rate as those during the 102 months. Thus, the FW CMEs constitute a small fraction of the general population: 6.3% (cycle 23), 3.1% (cycle 24), and 4.6% (combined). The corresponding fractions of CMEs associated with DH type II bursts are: 4.2% (cycle 23), 2.2% (cycle 24), and 3.1% (combined). Thus, CMEs associated with DH type II bursts constitute a small but special population of shock-driving CMEs. We also find that the fraction of CMEs producing DH type II bursts decreased drastically in cycle 24 following the trend of FW CMEs.

## 3. Statistical Properties

In this section, we obtain the statistical properties of CMEs associated with the DH type II bursts. Since Wind and SOHO spacecraft were launched within a year apart, we have excellent overlap between the two sets of observations. Moreover, the region of the corona imaged by the SOHO coronagraphs is also sampled by the radio bursts, so we have spatial information to combine with the radial resolution provided by the radio observations. The large number of events also enable us to perform statistical analyses of the radio-burst and CME properties. The derived properties confirm the earlier results obtained using a small number of type II bursts (Gopalswamy et al. 2001) and extend the results for a much larger sample, spanning almost two solar cycles. In comparing the statistical properties between cycles 23 and 24, we use all events unlike in section 2.3.1, where we compared over corresponding epochs in the two cycles. We consider CME kinematics, type II ending frequencies, SEP association, and the solar sources of the associated CMEs. The number of events is different from the one used in Fig. 6 mainly because of SOHO/LASCO data gaps. In addition to the 4-month data gap in cycle 23, there are no LASCO data for 22 type II bursts. This resulted in 496 CMEs from 1996 May to 2017 September, which we use in obtaining properties of CMEs associated with DH type II bursts.

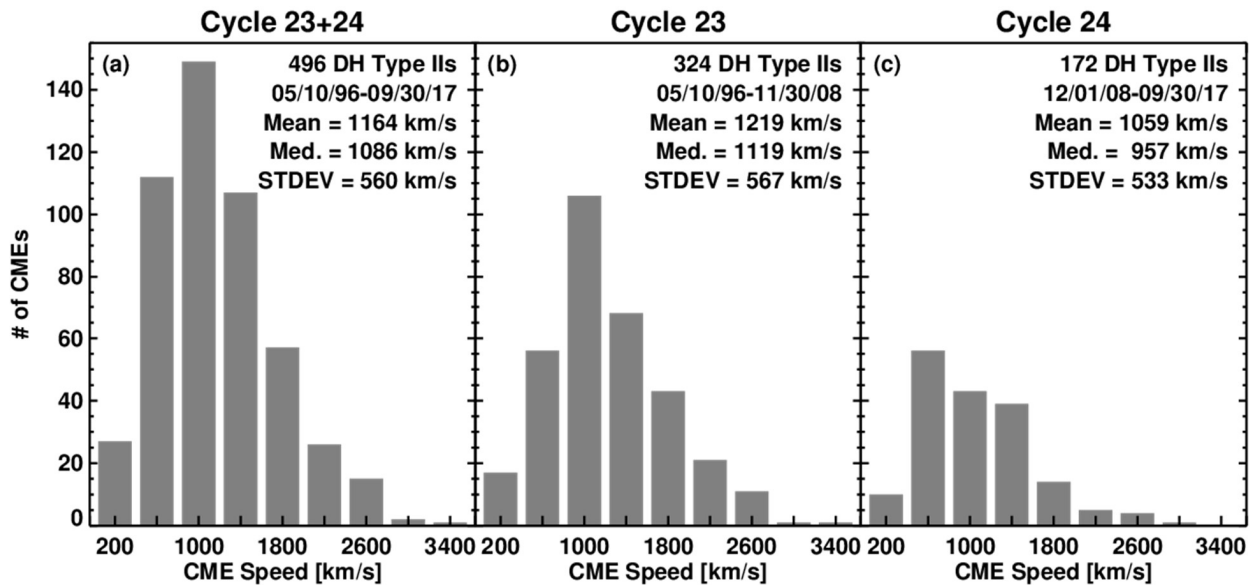

Figure 7. Speed distributions of CMEs associated with DH type II bursts. (a) all events from solar cycles 23 and 24, (b) cycle-23 events, and (c) cycle-24 events. Mean, median and standard deviation of the bursts are indicated on the plots, along with the number of events in each cycle. Unlike in section 2.3.1, all available CMEs in cycles 23 and 24 are used in this plot.

**3.1 CME Kinematics**

Figure 7 shows that the CMEs associated with DH type II bursts generally fast with an average speed exceeding 1000 km/s. This is true for both solar cycles 23 and 24. These speeds are about 3 times the speed of the general population of CMEs. The average speeds are slightly larger than the 960 km/s obtained for CMEs associated with the first 100 DH type II bursts in SC 23 (1997-2000, Gopalswamy et al. 2001);  they are also larger than the average speed (~960 km/s) of 28 CMEs associated with <1 MHz type II bursts observed by the International Sun-Earth Explorer-3 (ISEE-3) during the interval 1979-1983 (Cane, Sheeley, and Howard, 1987).  The smaller speeds in the previous studies maybe due to sample sizes, phase of the solar cycle, inclusion of purely kilometric type II bursts, and the field of view of the coronagraph using which the CMEs were observed. Note that the speeds in Fig. 7 are all sky-plane speeds, which are likely to be underestimates for disk events. In fact, most of the CMEs in the 200 km/s bins in Fig. 7 are from close to the disk center suggesting that their true speeds are likely to be higher; the true speeds need to be high enough to be super-Alfvenic.

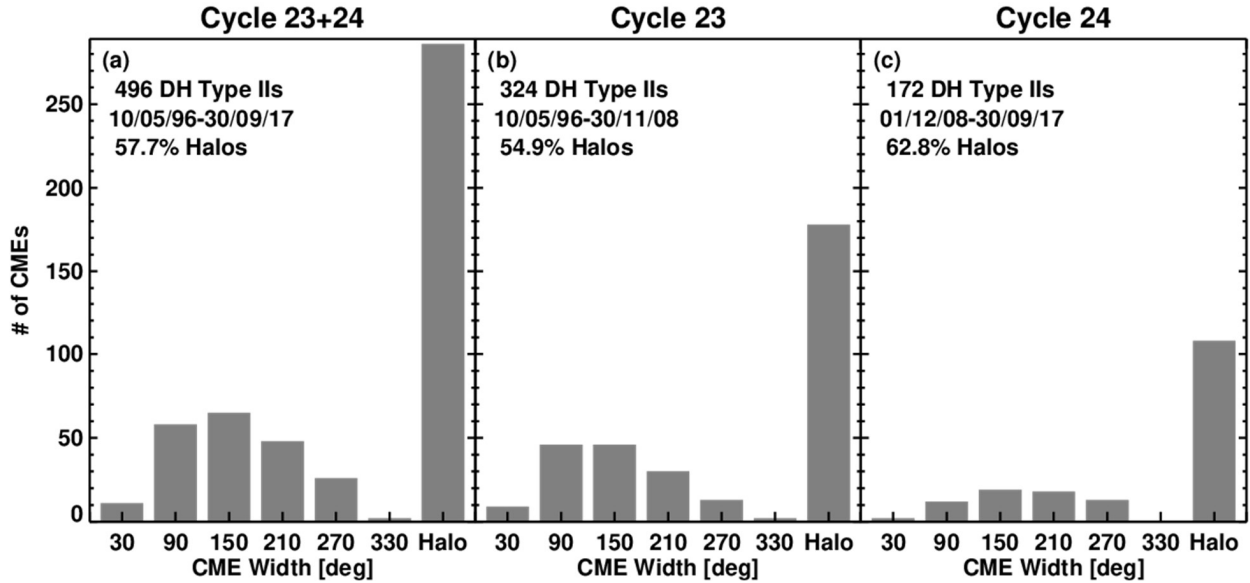

Figure 8. Width distributions of CMEs associated with DH type II bursts. (a) all events from solar cycles 23 and 24, (b) cycle-23 events, and (c) cycle-24 events. The number of events in each cycle and the fraction of halo CMEs in each population are noted on the plots.

The width distributions of CMEs associated with DH type II bursts are shown in Fig. 8. More than half of the CMEs are full halo CMEs (apparent width = 360º). This is a large halo fraction (~57%) compared to just ~3% in the general population of CMEs. The halo fraction in a CME population is a good indicator of the average energy of the population (Gopalswamy et al. 2010). Halo CMEs are inherently wide. This became clear when CMEs were observed quadrature by SOHO and STEREO. Halo CMEs observed by SOHO were viewed by one of the STEREO spacecraft in quadrature and found to be typically about 90º wide. We see in Fig. 8 that even the non-halo CMEs have widths >60º. Gopalswamy et al. (2001) referred to the CMEs producing DH type II bursts as FW CMEs suggesting that these CMEs are generally more energetic. This was also noted by Cane, Sheeley, and Howard (1987) for kilometric type II bursts. CMEs need to be fast (super-Alfvenic) to drive a shock but the shock's efficiency in accelerating particles depends on the CME width. Kahler et al. (2019) found that fast but narrow (width <60º) CMEs may not accelerate particles to significant levels: fast and narrow CMEs may drive a bow shock, whereas FW CMEs drive a piston driven shock with a significant shock sheath. Figure 8 shows another interesting result: the halo fraction is about 8% larger in cycle 24. This has been attributed to the weakened heliospheric total pressure, which allows CMEs to expand and appear as halos.

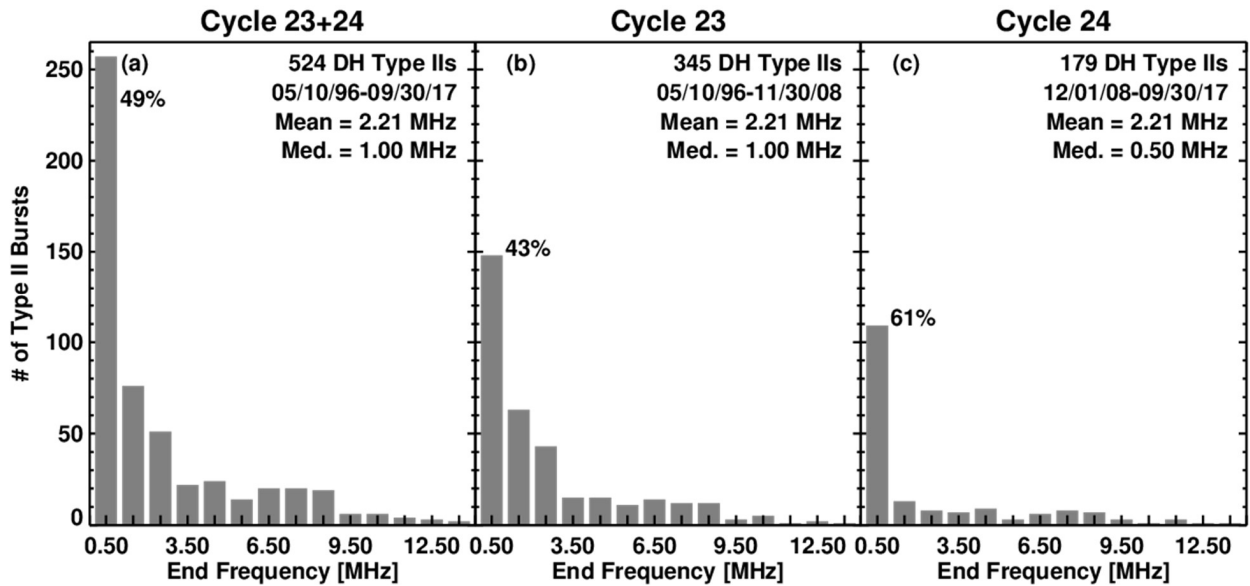

Figure 9. Distributions of ending frequencies of DH type II bursts. All the type II bursts in the catalog are used in (a) except for the 4 cases with uncertain frequencies. Cycle 23 and 24 events are given in (b) and (c), respectively. The number of events, interval of observation, mean, and median are shown for each distribution. The percentage of events in the 0.5 MHz bins are: 49% (a), 43% (b), and 61% (c).

## 3.2 Ending Frequencies of Type II Bursts

The ending frequency of type II bursts is a good indicator of the energy of the underlying CME. For example, type II bursts with emission components from metric to kilometric wavelengths have almost one-to-one correspondence with large SEP events (Gopalswamy 2006) because both the phenomena require strong shocks. In fact, all SEP events with ground level enhancement (GLE) are associated with such long-enduring type II bursts (Gopalswamy et al. 2012). It has recently been found that the frequency of type II emission at the end times of sustained gamma-ray emission (SGRE) from the Sun is ~200 kHz suggesting that the underlying shocks continue to accelerate high-energy protons to tens of Rs into the heliosphere (Gopalswamy et al. 2018a). It must be noted that there is a population of type II bursts starting and ending in the kilometric wavelengths. These are associated with relatively slow CMEs (~500 km/s) that continue to accelerate through the coronagraphic field of view and become super-Alfvenic at tens of Rs from the Sun (Gopalswamy 2006). On the contrary, CMEs associated with metric type II bursts are of moderate speed (~600 km/s) and become super-Alfvenic very close to the Sun and quickly decelerate in the IP medium (Gopalswamy et al. 2005a). The combined population of DH and DH-km type II bursts listed in this catalog are of much higher average speed (~1100 km/s) and hence energy because faster CMEs are generally more massive. Therefore, the low ending frequency of type II bursts is an indicator of an energetic CME. Fig. 9 shows the distribution of ending frequencies for cycles 23 and 24 along with that of the joint distribution. About half of the type II bursts end at frequencies below 0.5 MHz. The underlying CMEs are expected to be more energetic compared to the rest of the CMEs. Two-thirds of type II bursts reach at least until 1.5 MHz (the first two bins in the distributions). The fraction of type II bursts reaching 0.5 MHz is significantly higher in cycle 24 (61% compared to 43% in cycle 23). This can also be attributed to the weakened state of the heliosphere (the Alfven speed in particular, as described in Gopalswamy et al. 2014).

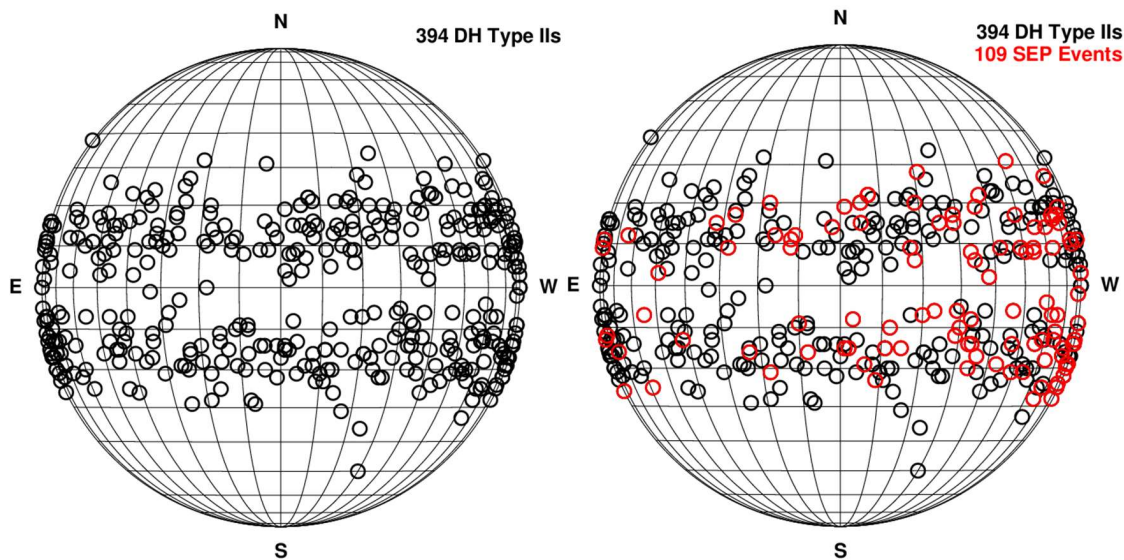

Figure 10. (left) Solar source locations of CMEs associated with frontside DH type bursts. (right) Same as left, but the sources of 109 CMEs associated with large SEP events are distinguished in red. Recall that out of the 524 DH type II bursts, CME information is available only for 496 bursts. Of the 496 bursts, only 394 (or~80%) have sources on the frontside of the Sun.

## 3.3 Solar Sources

Figure 10 shows the heliographic coordinates of the CMEs associated with DH type II bursts. Only 394 frontside sources (out of 524) or 80% are plotted. The solar sources are predominantly located in the latitude range of ±30º, corresponding to the active region belt. Only 11 of the 394 events (or 2.8%) originated outside of ±30º latitudes. DH type II bursts require FW CMEs that can only be powered by the high magnetic energy stored in active region magnetic fields. This is consistent with the overall correlation between SSN and the Type II burst occurrence rate shown in Fig. 6. Some DH type II bursts are associated with FW CMEs originating in non-spot (quiescent filament) regions. Even in these cases, the eruptions are from low latitudes (Gopalswamy et al. 2015) again for the reason of high magnetic energy stored.

## 3.4 SEP Association

Type II bursts are good indictors of CME-driven shocks because they accelerate electrons. The same shocks also accelerate protons and heavier ions that can be observed as SEP events in space when the solar source is magnetically well connected to the observer. It has been shown before that there is a high degree of association between DH type II bursts and large SEP events. Cliver, Kahler, and Reames (2004) reported that the degree of association between type II bursts and ~20 MeV SEP events with peak intensities of ≥10 pfu (protons $cm^{-2}$ $s^{-1}$ $sr^{-1}$ $MeV^{-1}$) is significantly higher when an event has both metric and DH components instead of just metric type II bursts. Gopalswamy et al. (2008b) reported a high degree of association between DH type II bursts originating from the western hemisphere and large SEP events (>10 MeV particle intensity >10 pfu). In fact, the association is 100% when the western hemispheric DH type II bursts are associated with very fast CMEs (speed >1800 km/s).

The source locations of DH type II bursts that are associated with large SEP events are shown in Fig. 10 (right). The latitude range of SEP sources is even tighter than that of the DH type II bursts. In fact, none of the 11 DH type II bursts occurring at latitudes >30º are associated with SEP events. Most of the SEP-associated type II bursts originate from the western hemisphere. Only 23 of the 109 SEP events (or 21%) originate from the eastern hemisphere. The CMEs underlying eastern SEP events are generally much faster and wider, so the extended shock surface crosses field lines connecting to Earth. In many cases, the shock becomes weak and the SEP intensity may not increase above the 10 pfu level to be counted as a large SEP event. Only 109 of the 394 DH type II bursts (27.7%) are associated with a large SEP event. The low rate of association can be attributed to several reasons. (i) Eastern events are not well connected to Earth. (ii) SEP events are not distinguished when type II bursts occur in quick succession, resulting in an undercount. (iii) Events from the western hemisphere have DH type II bursts that are associated with <10 pfu SEP events, and hence are not counted as large SEP events. If we consider only western hemispheric events, the degree of association is higher: 86/197 or 44%. The rate is much higher when minor SEP events (intensity <10 pfu, Chandra et al. 2013) are included. The association rate also depends on the CME speed: all western DH type II bursts with V>1800 km/s are associated with large SEP events (Gopalswamy et al. 2008b). It must be noted that SEP events are observed from an eastern source when it is magnetically connected to another vantage point (e.g., STEREO spacecraft trailing well behind Earth).

During the period considered in Fig. 10, there are 152 large SEP events, which means the sources of 43 large SEP events are not plotted in Fig. 10. Eight of the 43 SEP events occurred during LASCO data gaps; they are all associated with DH type II bursts, but their solar sources are not plotted. In 19 cases, the sources were behind the limb and associated with DH type II bursts. In 6 cases, the SEP event is an energetic storm particle event (ESP, see, e.g., Mäkelä et al. 2011) as confirmed by the presence of an IP shock at the time of the event. In the remaining 10 cases, we do not have a corresponding DH type II burst listed in the catalog. A closer examination shows that there are small fuzzy features in the radio dynamic spectra that may be extremely weak type II bursts. In three cases, there are type II bursts at frequencies below 1 MHz and hence are not listed in the catalog.

## 4. Discussion

As we noted before, the DH type II bursts identify a small group of energetic CMEs that drive shocks. The overlap in the spatial domain observed by the SOHO coronagraphs and Wind/WAVES instruments has been extremely useful in understanding shocks and particle acceleration by them. Here we point out three cases that demonstrate the power of combining radio and white light observations to investigate CMEs and shocks.

(1) When SOHO was disabled over a 3-month period in 1998, a GLE event occurred on 1998 August 24. As expected, the event was associated with a type II burst with emission components from metric to kilometric wavelengths (Reiner et al. 1999; Gopalswamy et al. 2012). Even though there were no white-light observations, the associated CME was observed in-situ by the Wind spacecraft, so it was possible to estimate the CME speed near the Sun to be in the range 1400 to 1700 km/s.

(2) Coronagraphs can track CMEs and their shocks up to ~30 Rs from the Sun. Type II bursts are observed at these distances and often beyond, so the evolution of CME shocks can be followed in the IP medium. Gopalswamy et al. (2018b) reported on the evolution of an IP shock by combining coronagraph observations, radio observations from Wind/WAVES, and in-situ observations from the Wind spacecraft. They were able to calibrate the shock speed obtained from the drift rate of the DH type II burst with the shock speed from coronagraph observations to obtain speeds in the IP medium beyond the coronagraph FOV.

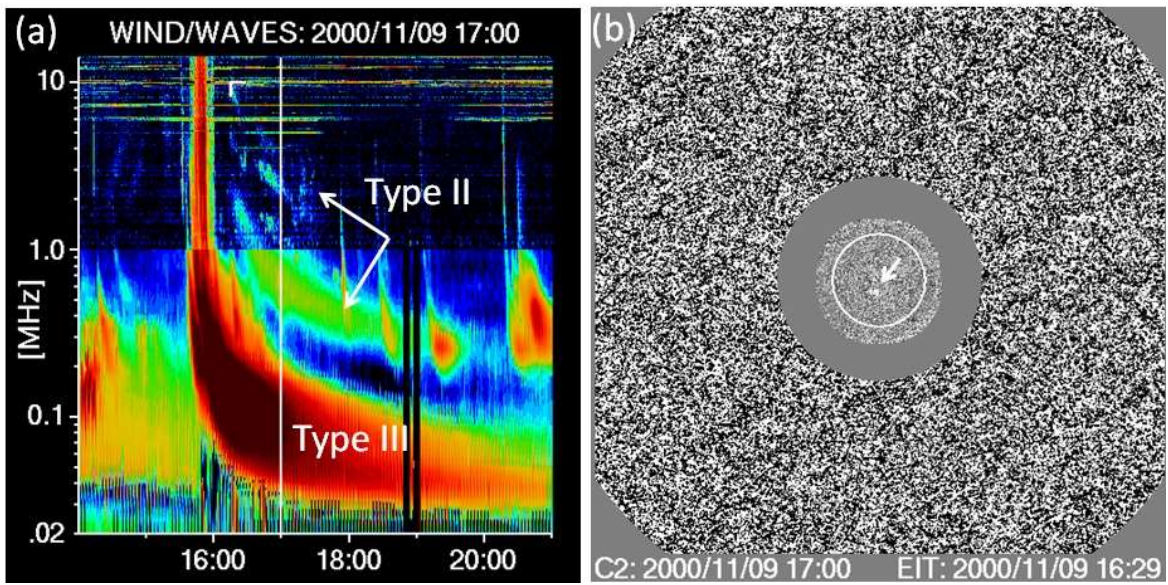

Figure 11. (a) Wind/WAVES dynamic spectrum showing the 2000 November 9 DH type II burst. (b) Coronagraph image taken during the type II burst with a SOHO/EIT difference image superposed. The "pepper and salt" appearance of the coronal image is due to the contamination of the coronal signal caused by the secondary particles produced when solar energetic particles hit the SOHO telescopes. This phenomenon is often referred to as a "snowstorm". The eruption from S11E10 is pointed at by an arrow. The vertical white line at 17:00 UT marks the time of the coronagraph image on the dynamic spectrum.

(3) When energetic CMEs occur in quick succession, a later CME can be obscured by the energetic particles hitting the coronagraph detector and saturating it. Figure 11 shows an intense DH type II burst, which occurred on 2000 November 9 and is listed in the catalog without an associated CME. The type II burst has multiple drifting lanes in the RAD2 frequency range, but a single intense burst in the RAD1 frequencies. The burst had possible emission components down to ~40 kHz, but not at the local plasma frequency at the Wind spacecraft. The solar source of the burst is identified as a disk center eruption (S11E10) accompanied by an M1.0 flare from AR 9221. However, the CME was not observed because the coronagraph detectors were saturated by energetic particles from another large eruption that occurred on the previous day at 23:06 UT. The preceding eruption was from the west limb (N10W77) accompanied by an intense DH type II burst and an SEP event. The preceding SEP event was so powerful that it even affected the observation of its own CME in the C3 FOV. When we played the LASCO/C3 movies of the November 9 event, there was a hint of some CME motion in the southwest direction but could not be measured. There was no metric type II burst listed in the online Solar Geophysical Data, which means the DH type II burst is the only strong indicator of the CME.

The shock responsible for the type II burst arrived at the Wind spacecraft on November 11 at 4:12 UT followed by an ICME. The shock transit time was ~36.5 hr from the flare onset at 15:45 UT on November 9. The shock speed at 1 AU was ~900 km/s (https://www.cfa.harvard.edu/shocks/wi_data/00186/wi_00186.html). The shock was running into a high-speed stream which had a very low density (~1.1 $cm^{-3}$). This is confirmed by the observations of the Thermal Noise Receiver (TNR) on Wind as shown in Fig. 12. The low density and high-speed wind ahead of the shock is consistent with an extended coronal present just to the west of the solar source region.

Taking the shock speed (900 km/s at 1 AU) to be the same as the driver speed, since the shock transit speed is 1134 km/s corresponding to the transit time of ~36.5 hr, and the 1-AU shock speed is 900 km/s, we can estimate initial speed of the CME at the Sun as 1368 km/s. The transit time T can also be used in the empirical shock arrival model, $T = AB^V + C$ with $A = 151.002$, $B=0.998625$, and $C = 12.5981$ (Gopalswamy et al. 2005b) to get the CME initial speed $V$ as 1340 km/s, consistent with the above estimate. Note that the derived speed is also consistent with the average speed of CMEs causing type II bursts in cycle 23 (Fig. 7b).

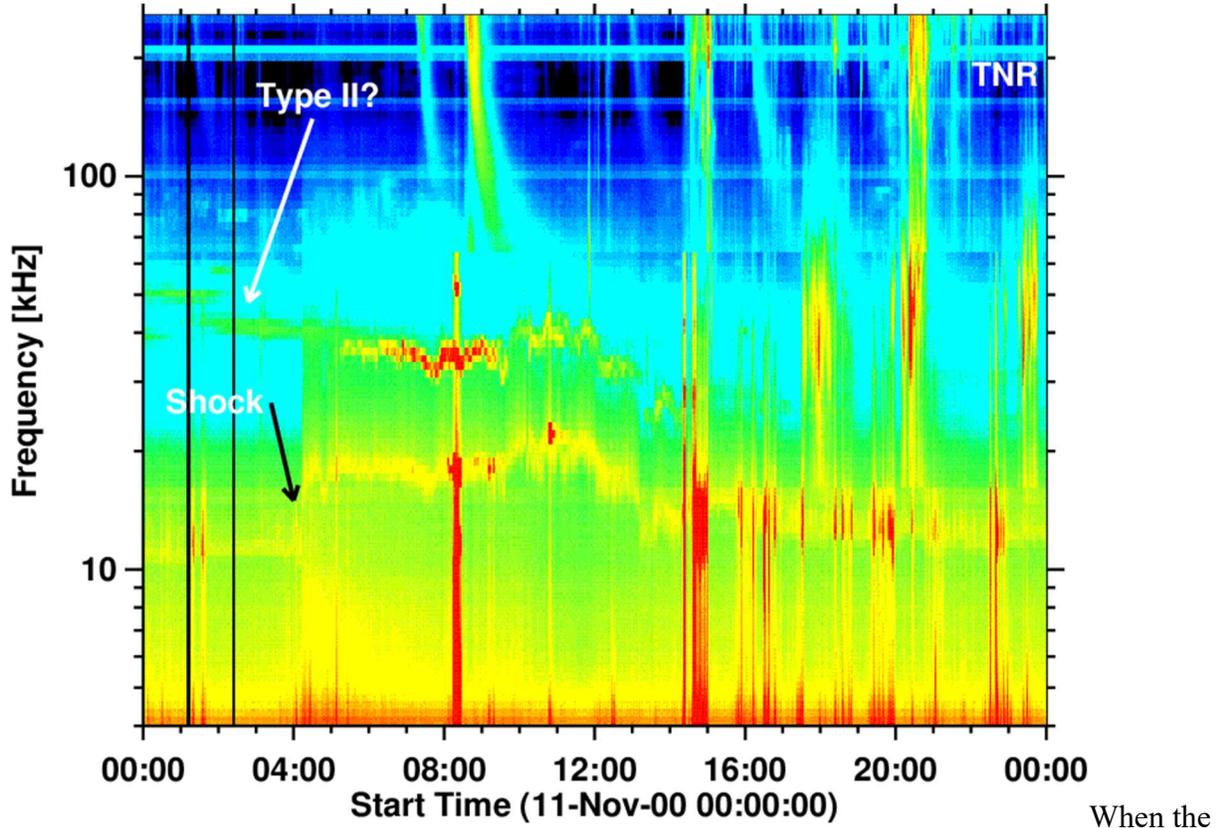

Figure 12. Wind/WAVES Thermal Noise Receiver (TNR) dynamic spectrum showing the shock arrival at the Wind spacecraft. The local plasma frequency is unusually low ~10 kHz, consistent with the low density observed by Wind's SWE instrument. A possible type II burst is observed at ~40 MHz, which is not at the local plasma frequency.

The shock speed can also be derived from the drift rate ($df/dt$) of the type II burst. When the type II burst is intense below 1 MHz, we measured $df/dt$ between 1 and 0.5 MHz as $2.08\times10^{-4}$ MHz s$^{-1}$. The mid frequency in this range is $f = 0.75$ MHz. The shock speed is given by eq. (1). The density scale height $L = r/2$ for density varying with heliocentric distance $r$ as $r^{-2}$ in the IP medium. Using the CME initial speed derived above, we can estimate the heliocentric distance of the shock/CME from eq. (1) as $r = Vf/(df/dt)$. With $V = 1340$ km/s, $f = 0.75$ MHz, $df/dt = 2.08\times10^{-4}$ MHz s$^{-1}$, we get $r \sim 7$ Rs. This distance is reasonable as we know from other events where both type II and CME observations are available.

## 5. Summary

We have described the online catalog consisting of all DH type II radio bursts manually identified from the Wind/WAVES dynamic spectra. The catalog also provides information on the associated CME, flare, solar source, and SEP event. The catalog has links to plots and movies that can be used to further investigate the underlying solar eruptions. The movies also have links to measurement tools that work on images and spectra associated with the type II bursts. It is possible that some weak type II bursts are missed in this catalog. When any of them are identified, the catalog will be updated by including such events. The solar activity is currently very low, so we have not observed any DH type II bursts for the past two years. However, the catalog will be maintained and updated periodically. We derived the statistical properties the bursts and the associated CMEs taking advantage of the >500 type II bursts. CMEs associated with DH type II bursts are sufficiently fast and wide that they drive shocks and hence accelerate particles. The Wind/WAVES observations span over two solar cycles, so we were able to study the solar cycle variation of the occurrence rate of the bursts. The main results derived from the catalog can be summarized as follows.

1. DH type II bursts isolate shock-driving energetic CMEs that constitute a small fraction (~3.1%) of all CMEs.

2. The occurrence rate of type II bursts follows the sunspot number but follows the number of fast and wide CMEs even more closely.

3. The average speed of CMEs associated with DH type II bursts is ~1164 km/s, which shows very little variation with solar cycle.

4. More than half of the CMEs associated with DH type II bursts are halos, indicating that these CMEs are more energetic on average. The halo fraction in DH type II bursts of solar cycle 24 is larger than that in cycle 23.

5. About half of the DH type II bursts end at frequencies below 0.5 MHz, suggesting that the underlying shocks are relatively strong.

6. The drift of DH type II bursts provide a simple way to track shocks in the IP medium because of the simple density variation with distance can be calibrated using white-light CME data.

7. Occasionally, DH type II bursts can be used to infer the underlying CMEs when white-light data are not available or when CMEs are obscured by the presence of SEPs from preceding CMEs.

**Acknowledgments**

This work benefited from NASA's open data policy. We thank the SOHO, STEREO, SDO, and Wind teams for making their data available on-line that are utilized in constructing this catalog. We also acknowledge the use of GOES X-ray and proton data made available online by NOAA. We acknowledge the use of version 2 sunspot data from the World Data Center - Sunspot Index and Long-term Solar Observations (WDC-SILSO), Royal Observatory of Belgium, Brussels. Work supported by NASA's LWS TR&T program. We thank the anonymous referee and the editor Susanna Finn for helpful comments.